\documentstyle[preprint,aps]{revtex}
\begin{document}
\draft              

\title{\bf Exact N - vortex solutions to the 
Ginzburg - Landau equations for $\kappa=1/\sqrt{2}$ }

\author{\bf A.V.Efanov}

\address{Institute of Semiconductor Physics, 630090, Novosibirsk, Russia}

\date{\today}
\maketitle

\begin{abstract}

        The N-vortex solutions to the two-dimensional Ginzburg - 
Landau equations for the $\kappa=1/\sqrt{2}$ parameter  are 
built. The exact solutions are derived for the vortices with 
large numbers of the magnetic flux quanta. The size of vortex 
core is supposed to be much greater than the magnetic field 
penetration depth. In this limiting case the problem is reduced 
to the determination of vortex core shape.  The corresponding 
nonlinear boundary problem is solved by means of the methods of 
the theory of analytic functions.

\end{abstract}


\narrowtext
\section{Introduction}

In different areas of physics there is a wide interest in
nonlinear field equations admitting particlelike solutions. The
Ginzburg-Landau (GL) theory \cite{L1} is a well known example of
this kind. In the case of a hard superconductor subjected to 
an external magnetic field the GL equations give vortex-line 
solutions. The properties of the vortex states in the GL theory 
are investigated in two limiting cases \cite{L2}. In the first 
case the sector of the fields near the lower critical field 
$H_{c_1}$ is considered.  In this sector a mixed state of the 
type-II superconductor is a system of widely separated vortices. 
In this case the problem is solved for the unit vortex. Due to 
the cylindrical symmetry of the system the corresponding GL 
equations are reduced to the set of the ordinary differential 
equations. In the second case the sector of the fields near the 
upper critical field $H_{c_2}$ is considered. In this region the 
periodical lattice of the vortices is the equilibrium state of 
the superconductor. Due to the small value of the order 
parameter the GL equations turn out to be linear. It allows one 
to investigate them by the usual methods.

In this work the exact solutions of nonlinear GL equations are
built. The solutions are derived in the special case of the
$\kappa =\delta/\xi= 1/\sqrt{2}$, where $\delta$ is the magnetic
field penetration depth, $\xi$ is the fluctuation radius of the
order parameter, and in the asymptotic of the large magnetic
fluxes of each vortex $\Phi \gg \Phi_0$, where $\Phi_0=\pi \hbar
c/e$ is the magnetic flux quantum. In the limiting case when the
vortex core sizes considerably exceed the typical lengths
$\delta$ or $\xi$ the problem is reduced to the purely
geometrical one, to the determination of the shape of the contour
confining the vortex cores. Such a problem proves to be 
nonlinear.  Nevertheless, it is solved by means of methods 
of the theory of analytic functions.

This case is interesting since the quantum properties of the 
superconductor are manifested in the classical limit. The 
$\kappa$ parameter determines the value of tension coefficient 
on the interphase of normal and superconducting phase. It is at 
$\kappa =1/\sqrt{2}$ that the coefficient vanishes\cite{L3} and 
causes the agreement of all three critical magnetic fields 
$H_{c_1}=H_{c_2}=H_c$.  Therefore, the material with such value 
of the $\kappa$ parameter possesses the properties of both 
type-I and type-II superconductors. On the one hand, the 
vortices with large magnetic fluxes are the incorporations of 
the normal phase. They should be considered as the parts of the 
intermediate state of type-I superconductor. On the other hand, 
such regions of the normal phase are nothing but the vortices. 
As shown below their shape depends on the coordinates of the 
zeros of the order parameter and the numbers of the magnetic 
flux quanta in every vortex, i.e. on the typical parameters of 
the mixed state.

The vortex states with large magnetic fluxes under consideration 
are not too "exotic" objects. Parallel with the macroscopic 
vortices the quantum vortices can also exist in superconductor.  
This is caused by the absence of the interaction among the 
vortices. The energy of the whole system depends only on the 
total magnetic flux \cite{L4}. At the given total magnetic flux 
the decay of the large vortices into small vortices does not 
lead to a gain in energy.

This consideration is based on the results of Ref.\cite{L4}. In
this work it is shown that the GL equations are reduced to a set
of the first order coupled equations for the vector potential and
the order parameter. By appropriate choice of the variables the
latter are written in the form of an equation similar to the
Poisson equation \cite{L5}. Its form is exactly the one that has
the asymptotically exact solutions \cite{L6}.

\section{General approach}

Let us consider the two-dimensional distributions of the complex
order parameter $\psi ({\bf r})$ and the vector potential ${\bf
A}({\bf r})$. Such distributions are realized for the
superconducting cylinder subjected to the longitudinal magnetic
field (see Fig. \ref{fig1}). If no edge effects are present all,
values depend on the transversal coordinates $x$ and $y$ only.

After rescaling the lengths and the fields the free energy of
superconductor per unit length of the sample is written as
follows
\begin{eqnarray}
\label{g8}
F & = & \int d^2 {\bf r} \left[ 
\frac{1}{2} {\bf B}^2+\frac{1}{2} \left| \left(  \nabla 
-i {\bf A} \right)  \psi \right| ^2  \right. \nonumber\\ 
 & & \left. \mbox{}+\frac{\lambda^2}{8} \left( \left|  
\psi \right| ^2-1 \right) ^2 \right], 
\end{eqnarray} 
where all the lengths are measured in the units of magnetic 
penetration depth $\delta$, and $\lambda =\sqrt{2}\kappa$ is the 
dimensionless coupling constant. For the chosen variables the 
physical magnetic field $B=H_c$ corresponds to the dimensionless 
field $B=\lambda/2$.

It is convenient to introduce the complex variables such as
\begin{equation}
\label{g9}
\begin{array}{cc}
z=x+i y, & \bar z=x-i y
\end{array}
\end{equation}
and the vector potential
\begin{equation}
\label{g10}
\begin{array}{cc}
A=\displaystyle \frac 12\left( A_x-i A_y\right) , &
\bar A=\displaystyle \frac 12\left( A_x+i A_y\right) .
\end{array}
\end{equation}
These notations us permit to present the equations derived in 
Ref.  \cite{L4,L5} for $\kappa=1/\sqrt{2}$ and, correspondingly, 
$\lambda =1$ in the most simple form. Such equations give the 
minimum of free energy (\ref {g8}). Unlike the GL equations they 
are the first order with respect the derivatives:
\begin{eqnarray}
\label{g14} 
\frac{\partial \psi 
}{\partial \bar z}-i\bar A \psi =0, \\ 
B+\frac 12\left( \left|  \psi \right| ^2-1\right) =0.  
\label{g15} 
\end{eqnarray} 
These equations give the vortexlike solutions with the magnetic
flux
\begin{equation}
\label{g16}
\Phi =\int d^2 r B=2\pi n,
\end{equation}
where $n$ is the positive integer number. The corresponding total
free energy of the system in dimensionless units turns out to be
equal exactly to $\pi n$.

The set of the coupled equations (\ref {g14}) and (\ref {g15}) is
reduced to one equation by the substitution of the form
\cite{L5}:
\begin{eqnarray} 
\label{g17} 
A & = & -\frac{i}{2} \frac{\partial}{\partial z} \left( u+2\ln |f| 
\right), \\ 
\label{g18} 
\psi & = & \frac f {\left| f \right| }\exp \left(-\frac{u}{2} 
\right), 
\end{eqnarray} 
where $u(x,y)$ is the real-valued function, and $f(z)$ is the 
analytic function. After the substitution the former equation 
of the system becomes the identity. The second equation takes 
the form 
\begin{eqnarray} 
\label{g19}
\Delta u & = & -4\pi \rho 
\left( u\right) -2\Delta \ln \left| f\right| \nonumber\\
 & = & -4\pi \rho \left( u\right) -4\pi \sum_{j=1}^Nn_j\delta 
({\bf  r- r}_j) , 
\end{eqnarray} 
where $\Delta $ is the Laplace two-dimensional operator,
\begin{equation}
\label{g22}
\rho (u)=\frac 1{4\pi }\left[ \exp(-u)-1 \right].
\end{equation}

Formula (\ref {g19}) differs from the analogous one of Ref.  
\cite{L5} in the last term of the right-hand side. Here 
singularities of the function $\ln \left| f(z) \right|$ are 
taken into account. The analytic function $f(z)$ vanishes like 
$\left( z-z_j\right) ^{n_j}$ at the positions of the vortices 
$z_j$, where $n_j$ is the positive integer number, $j=1,...,N$, 
and $N$ is the number of the vortices \cite{L5,L7}. This leads 
to the $\delta$ functions in the right-hand side of the equation 
(\ref {g19}).

The boundary condition for this equation is vanishing of the 
value $u(x,y)$ at the infinity. According to expression 
(\ref{g18}) this requirement gives equality $\left| \psi 
\right|=1$. The function $f(z)$ is to some degree an arbitrary 
one. Its form depends on the choice of the gauge of the vector 
potential.

The derived equations have the following physical meaning. They 
evidently describe the screening of the external positive point 
charges in a classical plasma. The value of the $u({\bf r)}$ is 
the electrostatic potential of the system of charges. The 
dependence of $\rho (u)$ gives the induced charge density 
described by the jellium model which consists of a negative 
smeared background and mobil charged fluid.

The exact solutions of nonlinear screening equation (\ref {g19})
are built in the limiting case $n_j\gg 1$. In this approximation
a sufficiently large positive potential is supposed to exist near
external point charges. As a consequence the screening results
from a complete repulsion of the positively charged fluid from
the regions surrounding the point charges. The induced charge
density turns out to be approximately constant within these
regions and is equal to $\rho \left( \infty \right) = -1/4\pi $.
The size of such region is much greater than the Debye length for
linear screening.

In these assumptions one can ignore the details of the potential
behavior on the scale of the order of the Debye length. Namely,
one can suppose that the system consists only of the uniformly
charged spatial region and a set of point charges. In this case 
the screening of the point charges can be achieved by means of 
selecting such a form of the region in which everywhere outside
the region both the potential and the electric field would
vanish. This problem is much easier than the initial one. Due to
its two-dimensionality it is solved by the methods of the theory
of functions of a complex variable \cite{L6}.

In the given case the results of the work \cite{L6} can be
rewritten in the following form. The solution is given by the
function
\begin{equation}
\label{g26} 
z=\omega (\zeta )=c_0+\sum_{j=1}^N\frac{c_j}{\zeta -\zeta _j}, 
\end{equation} 
which makes a conformal mapping of the inner region of the 
unit circle $\left| \zeta \right| \leq 1$ in an auxiliary plane 
of the complex variable $\zeta$ into the region of interest. The 
constants $\zeta _j$ obey the condition $\left| \zeta _j\right| 
>1$ for all $j$. The shape of the contour is specified 
parametrically: $z=\omega (\zeta )$ for $\zeta = \exp(i \alpha)$ 
and $0 \leq \alpha \leq 2\pi$.

The set of unknown constants $c_j$ and $\zeta _j$ is calculated
by means of the system of nonlinear algebraic equations
\begin{eqnarray}
\label{g28}
\omega \left( \xi _j\right) & = & z_j,  \\ 
\bar c_j\xi _j^2\omega ^{\prime }\left( \xi _j\right) & = & -4n_j,
\nonumber
\end{eqnarray}
where $\xi_j=1/{\bar \zeta}_j$, j=1,...,N; $\omega ^{\prime
}\left( \zeta \right)$ is the derivative of the function $\omega
\left( \zeta \right)$. Constant $c_0$ is determined from the
additional requirement that an inner point of the unit circle
should correspond to a given point within the uniformly charged
region.

The potential $u\left( x,y \right)$ is given by the expression
\begin{equation} 
\label{g29} 
u\left( x,y \right) =\frac 14 \left| z\right|^2-2 \mbox{Re} 
\, W(z), 
\end{equation} 
where
\begin{eqnarray}
\label{g30}
W(z)& =& \frac 18 v(z)+\sum_{j=1}^Nn_j\ln \left[ \frac{\bar \zeta
_j\zeta \left( z\right) -1}{\zeta \left( z\right) -\zeta _j}\right], 
\\ 
v(z)&=&\left| c_0\right| ^2+\sum_{j,k=1}^N\frac{ \bar c_jc_k}{1-\bar 
\zeta _j\zeta _k}+2\sum_{j=1}^N\frac{{\bar z}_jc_j}{\zeta \left( 
z\right) -\zeta _j}.
\nonumber
\end{eqnarray}
Here function $\zeta(z)$ is the inverse of $z=\omega (\zeta)$.

The results of the electrostatic problem allow one to write the
corresponding solutions of the GL equations easily. In the next
section the general formulas are given and some simple cases are
considered.

\section{Results and discussion}

Taking into account the fact that the gauge of the vector
potential can be an arbitrary one, let us take the function
$f(z)$ in the simpliest form
\begin{equation}
\label{g31} 
f(z) =\prod_{j=1}^N\left( z-z_j\right) ^{n_j}.  
\end{equation} 
Within the core of the vortices the order parameter and the vector 
potential will be described by the expressions:
\begin{eqnarray} 
\label{g32} 
\psi & = & \prod_{j=1}^N\left[ \frac{z-z_j}{\left| 
z-z_j\right| }\left| \frac{\bar \zeta _j\zeta \left( z\right) 
-1}{\zeta \left( z\right) -\zeta _j}\right| \right] ^{n_j} 
\nonumber\\ 
 & & \times \exp \left( -\frac{\left| z\right| ^2- \mbox{Re} \, 
v\left( z \right)}8\right),\\ 
A & = & -\frac i8\left[ \bar z-\eta 
(z)+4\sum_{j=1}^N\frac{n_j}{z-z_j} \right], 
\label{g35} 
\end{eqnarray}
where
\begin{equation}
\label{gg35}
\eta (z)={\bar c}_0+\sum_{j=1}^N\frac{{\bar c}_j \zeta(z)}{1- 
{\bar \zeta}_j \zeta(z) }. 
\end{equation}
In formula (\ref{g35}) value $A$ will be regular near the
points $z=z_j$, because the singularities of the function $\eta 
(z)$ are eliminated by the third term in square brackets.

Outside the vortex cores the order parameter and the vector
potential are described by
\begin{eqnarray}
\label{g34} 
\psi & = & \prod_{j=1}^N\left( 
 \frac{z-z_j}{\left| z-z_j\right| }\right) ^{n_j},\\
A & = & -\frac i2\sum_{j=1}^N\frac{n_j}{z-z_j}.
\label{g36}
\end{eqnarray}
In this region the induction of magnetic field $B$ equals zero.
On the contrary, within the vortex core the induction is ${
B}=1/2$. The last value corresponds to the field $B=H_c$ measured
in the physical units.

Formulas (\ref{g32})-(\ref{g36}) give the solutions only when 
the vortex cores are connected to each other \cite{L6}. For the 
sets of widely separated vortices these formulas should be 
applied to each connected group. These expressions do not work 
for the regions of "Swiss cheese" type. A solution of an 
electrostatic problem of this kind was obtained in Ref.  
\cite{L8}.

Formulas (\ref{g32})-(\ref{g36}) present the desired results
as the functions of the auxiliary variable $\zeta$. To obtain the
explicit dependence of these functions on the variable $z$ it is
necessary to invert the equality $z=\omega (\zeta)$ with respect
to the variable $\zeta$. It is possible only for some 
vortices.

Let us consider two special cases. The simpliest one is
the problem for the unit vortex in the origin of the coordinates
$z=0$. The corresponding solutions have the form
\begin{eqnarray}
\label{g40} 
\psi & = & \left( 
\frac{z}r\right) ^n\exp \left[ -\frac 18\left( \left| z \right| 
^2-r^2\right) \right] , \\ 
A & = & -\frac i8 \bar z, 
\nonumber 
\end{eqnarray} 
where $\left| z \right| \leq r$. The vortex core is a circle with
the radius $r=2\sqrt{n}$, where $n$ is the number of the flux
quanta of the magnetic field.

The problem for the twin vortices placed at the points $x=\pm a$
of the real axes can be regarded as the second example. The
vortex cores are overlapped if the distance $a$ obeys the
inequality $a \leq r$. Their shape is described by the conformal
mapping function
\begin{equation}
\label{g41}
z=\omega (\zeta )=\frac{2c_1\zeta }{\zeta _1^2-\zeta ^2},
\end{equation}
where
\begin{eqnarray}
\label{g42}
c_1 & = & a\sqrt{(p^2-1)\left( p+\sqrt{p^2-1}\right) }, 
\nonumber\\ 
\zeta _1 & = & \sqrt{p+\sqrt{p^2-1}}, \nonumber\\ 
p & = & \left( {\displaystyle \frac  ra} \right) ^2.  
\end{eqnarray} 
The order parameter is given by the expression
\begin{eqnarray} 
\label{g43} 
\psi & = & \left( \frac{z^2-a^2}{a \left| r+\sqrt{z^2-a^2+r^2}
\right| }\right)^n \nonumber\\
 & & \times \exp \left[-\frac{\left| z\right| ^2}8+\frac 
a4 \mbox{Re}\left(\sqrt{z^2-a^2+r^2}\right) \right].  
\end{eqnarray}
The corresponding formula for the vector potential is too bulky
to be presented here.
 
Note that solutions (\ref{g32})-(\ref{g35}) may be rewritten
in a form similar to the eigenfunction of a charged particle in
the lowest Landau level:
\begin{equation}
\label{g44}
\psi = F(z) \exp \left( -\frac{\left| z\right| ^2}8\right),
\end{equation}
where $F(z)$ is an analytic function. Such a presentation is 
evident for the following physical reasons. Within the normal 
phase the magnetic-field induction is approximately constant and 
equal to the critical field $H_c$. The order parameter is 
negligibly small here as well. The GL equations linearized with 
respect to the smallness $\left| \psi \right| \ll 1$ really give 
such solutions.

To derive the presentation (\ref{g44}) let us take function
$f(z)$ in the form $f(z)=\exp \left( W(z) \right)$.
Correspondingly, formula (\ref{g29}) gives the equality
$F(z)=f(z)$. The vector potential takes the form $A = -i \bar
z/8$.

In conclusion, let us discuss the validity of the model. It was 
assumed above that the order parameter depends on two 
coordinates only. This distribution is not realized in fact.  
According to the theory of the intermediate state of 
superconductors the domains of a normal phase decay near the 
surface of a sample into the alternating layers of normal and 
superconducting phases \cite{L9}. Thus, there is a dependence on 
the third coordinate directed along the external magnetic field.  
In this connection the existence of the macroscopic vortices may 
be provided only by a special mechanism for the attraction of 
vortices.  The pinning of vortices near space inhomogeneities or 
free boundary of the sample can become such a mechanism. The 
solution of this problem is beyond the scope of this article.

\acknowledgments 
This work was supported by Volkswagen Stiftung 
Foundation Grant No. I/68553. The author is grateful to 
E.G.Batyev and M.V.Entin for useful discussions.

\begin{figure}
\caption{ The geometry of the system. The  shaded region of 
the picture shows a unit vortex in a cylindrical specimen of 
superconductor.} 
\label{fig1} 
\end{figure} 

\end{document}